\documentclass[twoside,11pt]{article}

\usepackage{blindtext}

\usepackage[abbrvbib, preprint]{jmlr2e}

\usepackage{lastpage}
\usepackage{amsmath}

\usepackage{booktabs}   
\usepackage{array}     
\usepackage{multirow}  
\usepackage{makecell}  
\usepackage{caption}  
\usepackage{xcolor}
\usepackage{xspace}
\definecolor{spdlearncolor}{HTML}{c4301e}
\newcommand{\spdlearn}{\href{https://spdlearn.org}{\textcolor{spdlearncolor}{\emph{SPDLearn}}}\xspace}
\newcommand{\spdmodule}[2]{\href{#1}{\textcolor{spdlearncolor}{\textbf{\texttt{#2}}}}}

\ShortHeadings{\href{https://spdlearn.org}{\textcolor{spdlearncolor}{https://spdlearn.org}}}{Aristimunha, Ju, Collas, Bouchard, Mian, Thirion, Chevallier, and Kobler}
\firstpageno{1}

\begin{document}

\title{SPD Learn: A Geometric Deep Learning Python Library for Neural Decoding Through Trivialization}

\author{\name Bruno~Aristimunha \email b.aristimunha@gmail.com \\
\addr A\&O, LISN, Université Paris-Saclay, CNRS, Inria TAU/NERV, France \\
\addr Yneuro, France  \\
\addr University of San Diego, EUA
\AND
\name Ce~Ju \email ce.ju@inria.fr \\
\addr Inria, CEA, Université Paris-Saclay, Palaiseau, France
\AND
\name Antoine~Collas \email antoine.collas@inria.fr \\
\addr Inria, CEA, Université Paris-Saclay, Palaiseau, France
\AND
\name Florent~Bouchard \email florent.bouchard@centralesupelec.fr \\
\addr Université Paris-Saclay, CNRS, CentraleSupélec, L2S, France
\AND
\name Ammar Mian \email ammar.mian@univ-smb.fr \\
\addr Université Savoie Mont Blanc, LISTIC, France
\AND
\name Bertrand~Thirion \email bertrand.thirion@inria.fr \\
\addr Inria, CEA, Université Paris-Saclay, Palaiseau, France
\AND
\name Sylvain~Chevallier \email sylvain.a.chevallier@inria.fr \\
\addr A\&O, LISN, Université Paris-Saclay, CNRS, Inria TAU, France
\AND
\name Reinmar~Kobler \email reinmar.kobler@atr.jp \\
\addr Advanced Telecommunications Research Institute International, Kyoto, and RIKEN Artificial Intelligence Project, Tokyo, Japan.
\begin{center}
%{\small $^{\dagger}$ Equal Contributions}
\end{center}
}

%\thanks{Equal Contributions}

%\editor{My editor}

\maketitle

\begin{abstract}%   <- trailing '%' for backward compatibility of .sty file

\noindent Implementations of symmetric positive definite (SPD) matrix-based neural networks for neural decoding remain fragmented across research codebases and Python packages. Existing implementations often employ ad hoc handling of manifold constraints and non-unified training setups, which hinders reproducibility and integration into modern deep-learning workflows. To address this gap, we introduce \spdlearn, a unified and modular Python package for geometric deep learning with SPD matrices. \spdlearn provides core SPD operators and neural-network layers, including numerically stable spectral operators, and enforces Stiefel/SPD constraints via trivialization-based parameterizations. This design enables standard backpropagation and optimization in unconstrained Euclidean spaces while producing manifold-constrained parameters by construction. The package also offers reference implementations of representative SPDNet-based models and interfaces with widely used brain computer interface/neuroimaging toolkits and modern machine-learning libraries (e.g., MOABB, Braindecode, Nilearn, and SKADA), facilitating reproducible benchmarking and practical deployment.

\end{abstract}


\section{Introduction}

SPDNet provides a principled neural network framework for learning from covariance matrices representations by exploiting the geometry of the space of symmetric positive definite (SPD) matrices~\citep{huang2017riemannian}. By capturing second-order statistics while preserving geometric structure, SPDNet architectures have been successfully applied to EEG-based brain–computer interface studies~\citep{kobler2022spd,9805775,ju2025spd}, neuroimaging analysis~\citep{collas2025riemannian}, and related signal processing tasks, such as radar imaging~\citep{brooks2020deep}.

Despite their increasing use, existing SPDNet implementations remain fragmented across research codebases and Python packages. They often rely on ad hoc parameterizations to enforce manifold constraints, inconsistent implementations of spectral operators, or training pipelines that depend on specialized Riemannian optimizers, which complicates maintenance and reproducibility in practice.

To address these issues, we introduce \spdlearn, a unified and modular Python package for SPD matrix learning. A central design choice for manifold constraints is the systematic use of trivialization-based parameterizations~\citep{lezcano2019trivializations}: parameters constrained to the Stiefel or SPD manifolds are reparameterized in unconstrained Euclidean spaces through smooth mappings. The package also integrates with established BCI and neuroimaging toolchains, including \href{https://moabb.neurotechx.com/docs/index.html}{MOABB}~\citep{moabb2025} and \href{https://braindecode.org/stable/index.html}{Braindecode}~\citep{braindecode2025} for EEG benchmarking and \href{https://nilearn.github.io/stable/index.html}{Nilearn}~\citep{nilearn2026} for fMRI processing and analysis, as well as modern machine-learning domain adaptation libraries such as \href{https://github.com/scikit-adaptation/skada}{SKADA}~\citep{skada2024}, facilitating reproducible experimental pipelines.

%(e.g., householder decomposition-based or exponential/Cayley mappings via a skew-symmetric parametrization for Stiefel-manifold constrained \spdmodule{https://spdlearn.org/generated/bilinear/spd_learn.modules.BiMap.html}{BiMap} layer weights, and eigenvalue mappings such as exponential or softplus for SPD-manifold constraints)

\section{Design Principles Overview}

\begin{figure}[!ht]
    \centering
    \includegraphics[width=\linewidth]{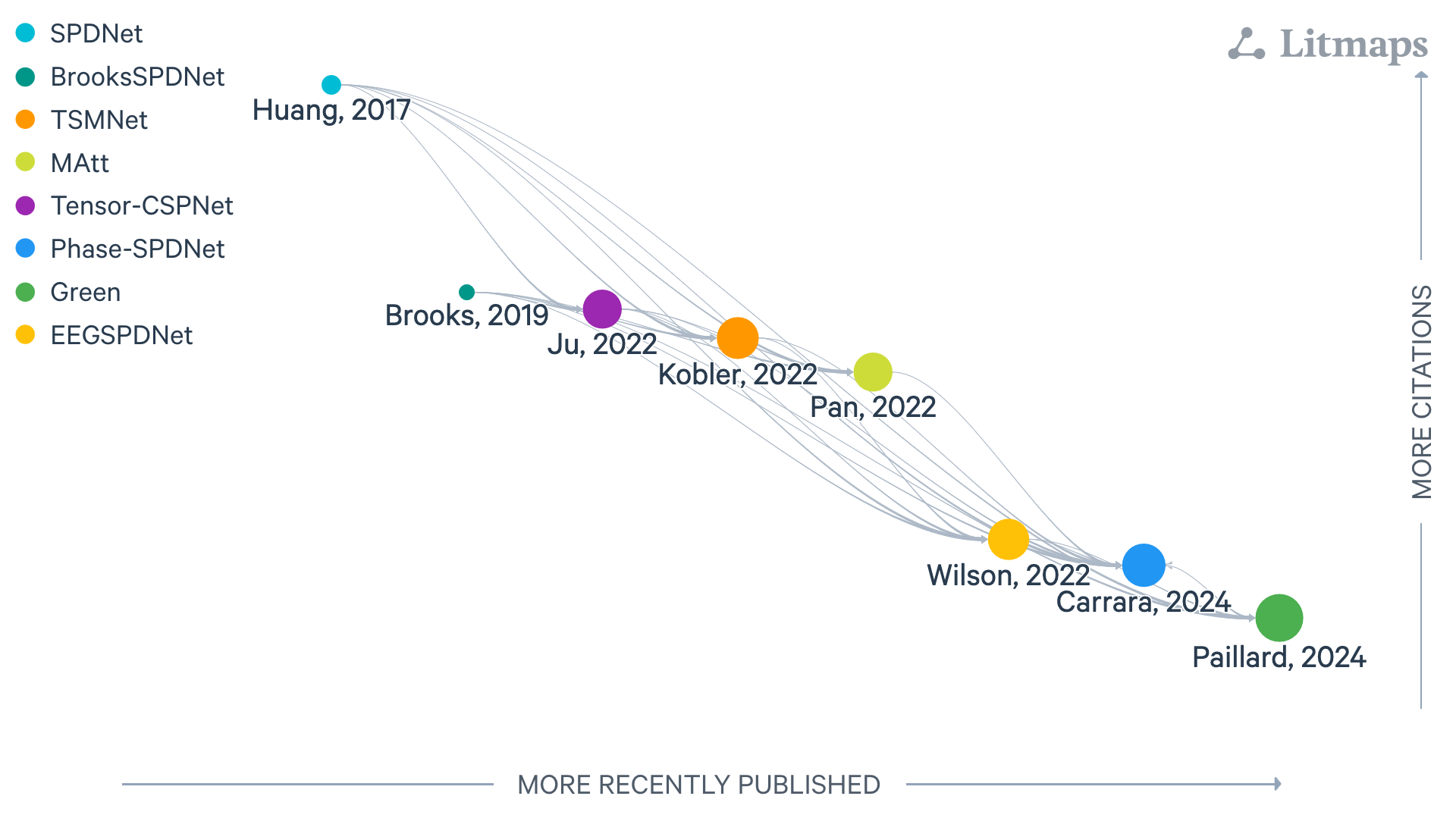}
    \caption{Literature Map: The x-axis denotes the released/publication date (more recent papers to the right) and the y-axis denotes citation count (more highly cited papers at the top). Directed edges indicate influence relationships, defined by citation links between papers. The map was generated with \href{https://www.litmaps.com/}{Litmaps} using data available as of January~2026.}
    \label{fig:litmap}
\end{figure}

The \spdlearn package is designed to be modular, extensible, and reproducible, organized into three components:
\begin{enumerate}\setlength{\itemsep}{2pt}\setlength{\parskip}{0pt}
    \item \texttt{functional}, which implements low-level operators for SPD matrices. It serves as the computational backbone for spectral operators by providing explicit forward and backward passes derived from spectral differentiation rules.
    \item \texttt{module}, which builds on \texttt{functional} to provide core neural-network layers for SPD matrix learning. These modules can be flexibly assembled to reproduce and readily extend a wide range of SPDNet-based architectures.
    \item \texttt{model}, which provides reference implementations of representative SPDNet-based architectures built from the modules in \texttt{module}, as summarized in Literature Map~\ref{fig:litmap} and Table~\ref{tab:spd_models}.
\end{enumerate}

For completeness, we detail key spectral operators, functions, and classes in \texttt{functional} and \texttt{module} in the Appendix.

\begin{table}[!h]t
\centering
\caption{Neural decoding models from \texttt{model} supported by \spdlearn.}
\label{tab:spd_models}
\begin{tabular}{p{6.8cm} p{8cm}}
\toprule
\textbf{Neural Decoding Model} & \textbf{Description} \\
\midrule
\textbf{Tensor-CSPNet}~\citep{9805775} 
& Captures discriminative information across temporal, spectral, and spatial domains separately using SPDNet for EEG motor imagery.\\

%\textbf{Graph-CSPNet}~\citep{10255369} & A follow-up SPDNet-based architecture for EEG motor imagery classification that jointly captures discriminative information in both the time–frequency and spectral domains using a proposed graph convolutional networks for SPD matrix-valued data, submitted to arXiv on 25 Oct, 2022.\\

\textbf{TSMNet}~\citep{kobler2022spd} & Introduces a sample covariance matrix pooling layer and a batch norm layer \spdmodule{https://spdlearn.org/generated/batchnorm/spd_learn.modules.SPDBatchNormMeanVar.html}{SPDBatchNormMeanVar} followed by an SPDNet classifier for EEG motor imagery.
\\
\textbf{Matt}~\citep{pan2022matt} & Introduces a manifold-valued attention module to SPDNet under Log-Euclidean geometry.\\
\textbf{Green}~\citep{paillard2025green} & Uses learnable wavelet convolutional features from EEGs followed by an SPDNet classifier.\\
\textbf{EEGSPDNet}~\citep{wilson2025deep} 
& Introduces a channel-specific convolutional layer and a sample covariance matrix pooling layer followed by an SPDNet classifier for EEG motor imagery.\\
\textbf{Phase-SPDNet}~\citep{carrara2025geometric}
& Applies a phase-space embedding using time-delayed coordinates followed by an SPDNet classifier for EEG motor imagery.\\
\bottomrule
\end{tabular}
\end{table}

\subsection{Trivialization}

In the SPDNet architectures, several parameter groups are subject to geometric constraints. First, the weight matrices in \spdmodule{https://spdlearn.org/generated/bilinear/spd_learn.modules.BiMap.html}{BiMap} layers are constrained to the Stiefel manifold. Second, some parameters are constrained to the SPD manifold, such as the bias matrix in \spdmodule{https://spdlearn.org/generated/batchnorm/spd_learn.modules.SPDBatchNormMean.html}{SPDBatchNormMean} and the scale/bias matrices in \spdmodule{https://spdlearn.org/generated/batchnorm/spd_learn.modules.SPDBatchNormMeanVar.html}{SPDBatchNormMeanVar}. In addition, certain parameters are constrained to be positive (e.g., scalar scale factors). In this package, these constrained parameters are handled via \emph{trivialization} techniques~\citep{lezcano2019trivializations}. 
Specifically, we construct a smooth mapping $\Phi:\mathbb{R}^d \rightarrow \mathcal{M}$, where $\mathcal{M}$ denotes the target manifold (e.g., Stiefel or SPD manifolds). 
Rather than optimizing directly on $\mathcal{M}$, we optimize unconstrained Euclidean parameters and enforce the constraint through $\Phi$ by construction. 
Gradients are backpropagated through $\Phi$ via the chain rule, enabling standard Euclidean optimization while producing manifold-valued (or otherwise constrained) parameters. For completeness, we detail the corresponding trivialization-based parameterizations in Appendix.

\subsection{Tutorials for Neural Decoding Tasks}

\spdlearn makes it easy to run SPD matrix learning models on EEG or fMRI benchmarks with a unified, reproducible setup built around tools such as MOABB~\citep{moabb2025} and Braindecode~\citep{braindecode2025}, Nilearn~\citep{nilearn2026}, and SKADA~\citep{skada2024}, while keeping data handling, preprocessing, and evaluation protocols consistent. In the documentation, we provide step-by-step tutorials with downloadable notebooks showing how to build end-to-end EEG motor imagery classification pipelines using \spdlearn, covering all models listed in Table~\ref{fig:litmap}, as well as a tutorial for fMRI classification. We also include animated visualizations to help users understand how each SPD neural network layer transforms data on the SPD manifold. %In particular, the MOABB + Hydra benchmark example offers a fully reproducible workflow: it uses MOABB for dataset loading and evaluation, Hydra for flexible experiment configuration, and a unified cross-validation protocol to compare multiple \spdlearn models.

\section{Conclusions}

We presented a Python package for geometric deep learning with SPD matrices, addressing fragmentation and inconsistency in existing SPDNet implementations. By adopting trivialization-based parametrizations, the package enables stable and efficient optimization of manifold-valued parameters within standard gradient-based learning frameworks. Its modular design and integration with established BCI and neuroimaging toolkits support reproducible experimentation and practical deployment of SPD-based models. Future work will focus on extending the package to additional architectures and learning paradigms, as well as broadening its applicability to larger-scale and more diverse neuroimaging and signal processing tasks.

%\blindmathpaper

%Here is a citation \cite{chow:68}.

\acks{This work was performed using HPC resources from GENCI--IDRIS (Grant 2025-AD011014817R1) on the Jean Zay supercomputer.}

\vskip 0.2in
\bibliography{sample}

% Acknowledgements and Disclosure of Funding should go at the end, before appendices and references

% \acks{}

% Manual newpage inserted to improve layout of sample file - not
% needed in general before appendices/bibliography.

\newpage

\appendix

\section{\texttt{Functional}}

\subparagraph{Eigenvalue forward operator.}
Given a symmetric matrix $X \in \mathbb{R}^{n\times n}$ with eigendecomposition $X = U \Lambda U^{\top}$, the function \texttt{modeig\_forward} applies a scalar function $f(\cdot)$ to the eigenvalues, producing $f(X) = U f(\Lambda) U^{\top}$, where $f(\Lambda)$ denotes element-wise application of $f$ to the diagonal entries of $\Lambda$. The function returns the transformed matrix together with the original eigenvalues and eigenvectors, which are cached for backpropagation.

\subparagraph{Eigenvalue backward operator.}
The function \texttt{modeig\_backward} implements the gradient of a spectral matrix
function $f(X)=U f(\Lambda) U^{\top}$ using the Loewner matrix formulation~\citep{ionescu2015matrix}.
Given the gradient $\nabla_Y \mathcal{L}$ with respect to the output
$Y=f(X)$, the gradient with respect to the input is computed as
\(
\nabla_X \mathcal{L}
=
U
\left[
L \odot
\left(
U^{\top} \, \mathrm{sym}(\nabla_Y \mathcal{L}) \, U
\right)
\right]
U^{\top},
\)
where $\odot$ denotes the Hadamard product and $L$ is the Loewner matrix associated
with $f$, defined element-wise by
\[ L_{ij} = \begin{cases} \dfrac{f(\lambda_i)-f(\lambda_j)}{\lambda_i-\lambda_j}, & i \neq j, \\ f'(\lambda_i), & i = j. \end{cases} 
\]

In practice, when eigenvalues are numerically close, the diagonal derivative $f'(\lambda)$ is used to ensure numerical stability and symmetry of the gradient. This formulation yields stable and theoretically consistent gradients for
eigenvalue-based matrix functions.

\subparagraph{Derived spectral operators.}

Based on this generic formulation, the following operators are implemented: $\mathrm{Log}(X) = U \log(\Lambda) U^{\top}$ (matrix logarithm), $\mathrm{Exp}(X) = U \exp(\Lambda) U^{\top}$ (matrix exponential), $|X| = U |\Lambda| U^{\top}$ (absolute eigenvalue mapping), $\mathrm{Clamp}(X) = U \max(\Lambda,\varepsilon I) U^{\top}$ (eigenvalue rectification, $\varepsilon > 0$), and $X^{\alpha} = U \Lambda^{\alpha} U^{\top}$ (matrix power, especially matrix square root and inverse square root are special cases with $\alpha=1/2$ and $\alpha=-1/2$).

\section{\texttt{Module}}

The layers \spdmodule{https://spdlearn.org/generated/bilinear/spd_learn.modules.BiMap.html}{BiMap}, \spdmodule{https://spdlearn.org/generated/modeig/spd_learn.modules.ReEig.html}{ReEig}, \spdmodule{https://spdlearn.org/generated/modeig/spd_learn.modules.LogEig.html}{LogEig}, and \spdmodule{https://spdlearn.org/generated/modeig/spd_learn.modules.ExpEig.html}{ExpEig} were introduced in~\cite{huang2017riemannian}. \spdmodule{https://spdlearn.org/generated/batchnorm/spd_learn.modules.SPDBatchNormMean.html}{SPDBatchNormMean} was introduced in~\cite{brooks2020deep}. \spdmodule{https://spdlearn.org/generated/bilinear/spd_learn.modules.BiMapIncreaseDim.html}{BiMapIncreaseDim} was introduced in~\cite{9805775}, and \spdmodule{https://spdlearn.org/generated/covariance/spd_learn.modules.CovLayer.html}{CovLayer} and \spdmodule{https://spdlearn.org/generated/batchnorm/spd_learn.modules.SPDBatchNormMeanVar.html}{SPDBatchNormMeanVar} were introduced in~\cite{kobler2022spd}.
\spdmodule{https://spdlearn.org/generated/dropout/spd_learn.modules.SPDDropout.html}{SPDDropout} was introduced in~\cite{wilson2025deep}.

\begin{sloppypar}
\begin{enumerate}
    \item Class of \spdmodule{https://spdlearn.org/generated/covariance/spd_learn.modules.CovLayer.html}{CovLayer}: This class computes the SPD representation $\Sigma = \mathcal{C}(X)$ for multivariate time series $X$, where $\mathcal{C}(\cdot)$ denotes a user-specified covariance operator. Supported operators include the estimator functions \texttt{covariance} (e.g., the population biased covariance matrix), \texttt{sample\_covariance} (e.g., sample (unbiased) covariance matrix), \texttt{real\_covariance} (e.g., the real part of the covariance (Gram) matrix for complex-valued time-series.), and \texttt{cross\_covariance} (e.g., a real-valued cross-frequency covariance matrix for wavelet-transformed time series).
    
    \item Classes of \spdmodule{https://spdlearn.org/generated/bilinear/spd_learn.modules.BiMap.html}{BiMap} and \spdmodule{https://spdlearn.org/generated/bilinear/spd_learn.modules.BiMapIncreaseDim.html}{BiMapIncreaseDim}: 
    Given an SPD input $X\in\mathbb{R}^{n\times n}$, \spdmodule{https://spdlearn.org/generated/bilinear/spd_learn.modules.BiMap.html}{BiMap} applies a bilinear congruence mapping $Y = W^{\top} X W$, where $W \in \mathbb{R}^{n\times m} (m\leq n)$ is a learnable orthogonal matrix enforced by a Stiefel constraint $W^{\top}W=I_m$. The weight matrix can be initialized using several strategies, including Kaiming uniform, orthogonal, and Stiefel initialization. When $m>n$, \spdmodule{https://spdlearn.org/generated/bilinear/spd_learn.modules.BiMapIncreaseDim.html}{BiMapIncreaseDim} embeds an SPD matrix $X\in\mathbb{S}_{++}^{n}$ into a higher-dimensional SPD matrix $\tilde{X}\in\mathbb{S}_{++}^{m}$ via
    \[
    \tilde{X} \;=\; A X A^{\top} \;+\; P,
    \]
    where $A\in\mathbb{R}^{m\times n}$ is a fixed embedding matrix given by the first $n$ columns of $I_m$, and $P\in\mathbb{R}^{m\times m}$ is a diagonal padding matrix defined by
    \[
    P_{ii}=
    \begin{cases}
    0, & i\le n,\\
    1, & i>n.
    \end{cases}
    \]
    Equivalently, this operation places $X$ in the top-left $n\times n$ block and pads the remaining dimensions with an identity matrix. This construction preserves positive definiteness and ensures $\tilde{X}\in\mathbb{S}_{++}^{m}$.

    \item Classes of \spdmodule{https://spdlearn.org/generated/modeig/spd_learn.modules.ReEig.html}{ReEig}, \spdmodule{https://spdlearn.org/generated/modeig/spd_learn.modules.LogEig.html}{LogEig}, and \spdmodule{https://spdlearn.org/generated/modeig/spd_learn.modules.ExpEig.html}{ExpEig}: Given a symmetric matrix $X=U\Lambda U^{\top}$, \spdmodule{https://spdlearn.org/generated/modeig/spd_learn.modules.ReEig.html}{ReEig} applies a rectification to the eigenvalues, $\mathrm{ReEig}(X)=U\,\max(\Lambda,\varepsilon I)\,U^{\top}$, to enforce numerical stability and positive definiteness. \spdmodule{https://spdlearn.org/generated/modeig/spd_learn.modules.LogEig.html}{LogEig} maps an SPD matrix to the symmetric Euclidean space via $\mathrm{LogEig}(X)=U\log(\Lambda)U^{\top}$, vectorizing the result by extracting the upper triangular entries optionally. Conversely, \spdmodule{https://spdlearn.org/generated/modeig/spd_learn.modules.ExpEig.html}{ExpEig} applies the inverse mapping $\mathrm{ExpEig}(X)=U\exp(\Lambda)U^{\top}$, to project symmetric matrices back onto the SPD manifold.

    \item Class of \spdmodule{https://spdlearn.org/generated/dropout/spd_learn.modules.SPDDropout.html}{SPDDropout}: \spdmodule{https://spdlearn.org/generated/dropout/spd_learn.modules.SPDDropout.html}{SPDDropout} randomly drops entire channels during training. Dropped channels are zeroed and their diagonal entries are replaced by a small constant $\varepsilon$, while the remaining channels are rescaled to preserve the expected value optionally, ensuring that the output remains positive definite.

    \item Classes of \spdmodule{https://spdlearn.org/generated/regularization/spd_learn.modules.TraceNorm.html}{TraceNorm} and \spdmodule{https://spdlearn.org/generated/regularization/spd_learn.modules.Shrinkage.html}{Shrinkage}: These regularization layers help improve numerical conditioning of SPD matrices. \spdmodule{https://spdlearn.org/generated/regularization/spd_learn.modules.TraceNorm.html}{TraceNorm} normalizes an SPD matrix by its trace, and \spdmodule{https://spdlearn.org/generated/regularization/spd_learn.modules.Shrinkage.html}{Shrinkage} applies Ledoit--Wolf-type shrinkage toward a scaled identity matrix.

    \item Class of \spdmodule{https://spdlearn.org/generated/residual/spd_learn.modules.LogEuclideanResidual.html}{LogEuclideanResidual}: This layer implements a residual connection in the Log-Euclidean framework, enabling skip connections between SPD-valued layers while preserving the manifold structure.

    \item Classes of \spdmodule{https://spdlearn.org/generated/batchnorm/spd_learn.modules.SPDBatchNormMean.html}{SPDBatchNormMean} and \spdmodule{https://spdlearn.org/generated/batchnorm/spd_learn.modules.SPDBatchNormMeanVar.html}{SPDBatchNormMeanVar}: Given a batch of SPD matrices $\{P_i\}_{i=1}^N$, \spdmodule{https://spdlearn.org/generated/batchnorm/spd_learn.modules.SPDBatchNormMean.html}{SPDBatchNormMean} estimates the batch Fr\'echet mean $\mathcal{G}$ under the affine-invariant Riemannian metric using a small number of Karcher flow iterations, and centers each input via the congruence transformation $\tilde{P}_i = \mathcal{G}^{-\frac12} P_i \mathcal{G}^{-\frac12}$. An optional learnable SPD bias matrix $B$ is then applied as $\hat{P}_i = B^{\frac12} \tilde{P}_i B^{\frac12}$. \spdmodule{https://spdlearn.org/generated/batchnorm/spd_learn.modules.SPDBatchNormMeanVar.html}{SPDBatchNormMeanVar} extends \spdmodule{https://spdlearn.org/generated/batchnorm/spd_learn.modules.SPDBatchNormMean.html}{SPDBatchNormMean} by additionally normalizing the dispersion of the batch. After centering at the running Fr\'echet mean $\mathcal{G}$, the matrices are whitened via a power transformation $\tilde{P}_i = \left(\mathcal{G}^{-\frac12} P_i \mathcal{G}^{-\frac12}\right)^{\alpha}$, $\alpha= w/\sqrt{\sigma^2 + \varepsilon}$, where $\sigma^2$ denotes the estimated Fr\'echet variance and $w$ is a learnable scaling parameter. An optional SPD bias is then applied by congruence.

    \item Class of \spdmodule{https://spdlearn.org/generated/batchnorm/spd_learn.modules.BatchReNorm.html}{BatchReNorm}: \spdmodule{https://spdlearn.org/generated/batchnorm/spd_learn.modules.BatchReNorm.html}{BatchReNorm} operates in a Euclidean (e.g., Log-Euclidean) representation. Given vectorized inputs $\{x_i\}_{i=1}^N$, the layer subtracts the batch or running mean $\tilde{x}_i = x_i - \mu$, and optionally applies a learnable bias, without enforcing SPD constraints.

\end{enumerate}
\end{sloppypar}

\section{Manifold-Valued Parameters}

\paragraph{Stiefel manifold-valued parameters.}
The weight matrix $W \in \mathbb{R}^{n \times m} \, (n>m)$ is required to have orthonormal columns $W^\top W = I_m$, which characterizes the Stiefel manifold $\mathrm{St}(n,m)$.
We parametrize $W$ using a trivialization induced by the geometry of the orthogonal group.
It is implemented by \texttt{torch.nn.utils.parametrizations.orthogonal}.
Three different trivializations are available.
The default one for real rectangular matrices is based on the Householder decomposition.
In this case, the weight $W$ is parametrized by $m$ Householder vectors $v_1,\dots,v_m$ such that $W$ corresponds to the $m$ first columns of $H_1\dots H_m$, where $H_i=I_n - 2 v_iv_i^\top / v_i^\top v_i$.
This way, each $H_i$ is orthogonal.
Moreover, each $v_i$ has zeros in positions $\{i+1,\dots,m\}$, ensuring the optimal number of degrees of freedom, \textit{i.e.}, $nm-m(m-1)/2$, and a mapping $\Phi$ from $\mathbb{R}^{nm-m(m-1)/2}$ onto $\mathrm{St}(n,m)$.
The two other trivialization options are a bit less efficient.
An unconstrained square matrix $X\in\mathbb{R}^{n\times n}$ is first transformed into a skew-symmetric matrix $A = (X-X^\top)/2$.
A $n\times n$ square orthogonal matrix is then obtained by applying either i). the Lie-group exponential map $\exp(A)$ (matrix exponential) or ii). the Cayley map $\left(I_n + 1/2A\right)\left(I_n - 1/2A\right)^{-1}$.
The weight matrix $W$ is finally obtained by taking the $m$ first columns of this square orthogonal matrix.
Hence, in that case, the mapping $\Phi$ is from $\mathcal{A}_n$ ($n\times n$ skew-symmetric matrices) onto $\mathrm{St}(n,m)$ and we have $n(n-1)/2$ degrees of freedom, which is sub-optimal.

\paragraph{SPD manifold-valued parameters.}
In batch normalization layers, the bias parameter needs to be an SPD matrix.
To deal with this constraint, a trivialization of the SPD manifold is employed.
Two different trivializations have been implemented in the \spdmodule{https://spdlearn.org/generated/manifold/spd_learn.modules.SymmetricPositiveDefinite.html}{SymmetricPositiveDefinite} parametrization.
In both cases, the mapping $\Phi$ is from $\mathcal{S}_n$ ($n\times n$ symmetric matrices) onto $\mathcal{S}^{++}_n$, thus featuring the optimal degrees of freedom, \textit{i.e.}, $n(n+1)/2$.
An unconstrained symmetric matrix $S\in\mathcal{S}_n$ is transformed into an SPD matrix by applying a mapping on its eigenvalues.
Given $S=U\Lambda U^\top$, the SPD matrix is obtained through $U f(\Lambda) U^\top$, where $f:\mathbb{R}\to\mathbb{R}^*_+$.
Two different mappings $f$ are available: i). the exponential and ii). the softplus functions.

\paragraph{Positive definite scalar parameters.}
In batch normalization layers, there is also a scale parameter, which has to be a positive definite scalar.
To ensure this constraint, the \spdmodule{https://spdlearn.org/generated/manifold/spd_learn.modules.PositiveDefiniteScalar.html}{PositiveDefiniteScalar} parametrization relies on a trivialization $\Phi$ from $\mathbb{R}$ onto $\mathbb{R}^*_+$.
Two options are available: i). the exponential and ii). the softplus functions.

\end{document}